\documentclass[floatfix, preprint,showpacs, showkeys, preprintnumbers, nofootinbib, superscriptaddress]{revtex4-1}
\usepackage{natbib}
\usepackage{times}
\usepackage{amssymb,amsbsy,amsmath,amsfonts}
\usepackage{bm}
\usepackage{graphicx}
\usepackage{float}
\usepackage{color}
\usepackage{xcolor}
\usepackage{morefloats}
\usepackage{rotating}
\usepackage{srcltx}
\usepackage{slashed}
\usepackage{subfigure}
\usepackage{multirow}
\usepackage{verbatim}
\usepackage{hyperref}
\usepackage{txfonts}
\usepackage{tabularx}
\usepackage{epstopdf}

\graphicspath{{./}{./img/}{./fig/}{./image/}{./figure/}{./picture/}}

\begin{document}

\title{Decuplet baryon masses in covariant baryon chiral perturbation theory}

\author{Xiu-Lei Ren}
\affiliation{School of Physics and
Nuclear Energy Engineering  and International Research Center for Nuclei and Particles in the Cosmos, Beihang University, Beijing 100191, China}

\author{Li-Sheng Geng}
\email[E-mail me at: ]{lisheng.geng@buaa.edu.cn}
\affiliation{School of Physics and
Nuclear Energy Engineering and International Research Center for Nuclei and Particles in the Cosmos, Beihang University, Beijing 100191, China}
\affiliation{Physik Department, Technische Universit\" at M\"unchen, D-85747 Garching, Germany}

\author{Jie Meng}
\affiliation{School of Physics and
Nuclear Energy Engineering and International Research Center for Nuclei and Particles in the Cosmos, Beihang University, Beijing 100191, China}
\affiliation{State Key Laboratory of Nuclear Physics and Technology, School of Physics, Peking University, Beijing 100871, China}
\affiliation{Department of Physics, University of Stellenbosch, Stellenbosch 7602, South Africa}

\begin{abstract}
  We present an analysis of the lowest-lying decuplet baryon masses in the
  covariant baryon chiral perturbation theory with the extended-on-mass-shell scheme up to
  next-to-next-to-next-to-leading order.
  In order to determine the $14$ low-energy constants, we perform a simultaneous fit of  the $n_f=2+1$ lattice QCD data from the PACS-CS, QCDSF-UKQCD, and HSC Collaborations, taking finite-volume corrections into account  self-consistently.
  We show that up to next-to-next-to-next-to-leading order one can achieve a good description of the lattice QCD and experimental data. Surprisingly, we note that the current lattice decuplet baryon masses can be fitted rather well by the next-to-leading order baryon chiral perturbation theory, which, however, misses the experimental data a little bit.
  Furthermore, we predict the pion- and strangeness-sigma terms  of the decuplet baryons by use of the Feynman-Hellmann theorem.
\end{abstract}


\date{\today}

\maketitle

\section{Introduction}

In the past few years many studies of the lowest-lying baryon octet and decuplet masses have been performed on the lattice~\cite{Durr:2008zz,Alexandrou:2009qu,Aoki:2008sm,
Aoki:2009ix,WalkerLoud:2008bp,Lin:2008pr, Bietenholz:2010jr,
Bietenholz:2011qq,Beane:2011pc} [see Ref.~\cite{Fodor:2012gf} for a comprehensive discussion of the various lattice chromodynamics (LQCD) simulations and the origin of their uncertainties].
These studies not only demonstrate the ability of LQCD
simulations to predict accurately  nonperturbative observables of the strong interactions, but also provide
valuable information that can be used to extract the
low-energy constants (LECs)  of  chiral perturbation theory (ChPT).
On the other hand, most LQCD calculations,  limited by the availability
of computational resources and the efficiency of algorithms~\cite{Beringer:1900zz},
still have to employ larger than physical light-quark
masses~$m_{u/d}$,~\footnote{It should be noted that
for a limited set of observables simulations with physical light-quark masses have  recently become available~\cite{Durr:2010aw,Bazavov:2012xda,Bazavov:2013vwa},} finite lattice volume~$V=TL^3$, and finite lattice spacing~$a$.
ChPT~\cite{Weinberg:1978kz,
Gasser:1983yg,Gasser:1984gg,Gasser:1987rb,Leutwyler:1994fi, Bernard:1995dp,Pich:1995bw,Ecker:1994gg,Pich:1998xt,Bernard:2006gx,Bernard:2007zu,Scherer:2012xha}
plays an important role in guiding  the necessary extrapolations to the physical world in  terms of light-quark masses~\cite{Leinweber:2003dg,Bernard:2003rp,
Procura:2003ig,Bernard:2005fy}, lattice volume~\cite{AliKhan:2003cu,Beane:2004tw}
and lattice spacing~\cite{Beane:2003xv,Arndt:2004we}, and in estimating the induced uncertainties.

As an effective field theory of low-energy QCD, ChPT has been rather successful in the mesonic sector, but the extension to the one-baryon sector turns out to be nontrivial. Because the baryon mass is not zero in the chiral limit, a systematic power counting is absent~\cite{Gasser:1987rb}. In order to restore the chiral power counting, the so-called heavy-baryon (HB) ChPT  was first proposed by Jenkins and Manohar~\cite{Jenkins:1990jv}. Although this approach provides a strict power counting, the heavy baryon expansion is nonrelativistic.  A naive application can lead to pathologies, e.g., in the calculation of the scalar form factor of the nucleon~\cite{Bernard:1995dp}.~\footnote{ This can be removed by resuming the leading kinetic operator to higher orders, equivalent to using the relativistic propagator~\cite{Becher:1999he}.} In addition, the HB ChPT is found to converge rather slowly in the three-flavor sector of $u$, $d$, and $s$ quarks. Later, covariant baryon chiral perturbation theory (BChPT) implementing a consistent power counting with different renormalization methods has been developed, such as the infrared (IR)~\cite{Becher:1999he} and the extended-on-mass-shell (EOMS) \cite{Gegelia:1999gf,Fuchs:2003qc} renormalization schemes.

In the past decades, the ground-state octet baryon masses have been studied rather extensively~\cite{Jenkins:1991ts,Bernard:1993nj,Banerjee:1994bk,Borasoy:1996bx,WalkerLoud:2004hf, Ellis:1999jt,Frink:2004ic,Frink:2005ru,Lehnhart:2004vi,Semke:2005sn,Semke:2007zz,
MartinCamalich:2010fp,Geng:2011wq,Young:2009zb,Semke:2011ez,Semke:2012gs,Lutz:2012mq,Bruns:2012eh,Ren:2012aj,
Ren:2013dzt}, especially, in combination with the $n_f=2+1$ LQCD data up to next-to-next-to-next-to-leading order (N$^3$LO)~\cite{Young:2009zb,MartinCamalich:2010fp, Geng:2011wq,Semke:2011ez,Semke:2012gs,Lutz:2012mq,Bruns:2012eh,Ren:2012aj,Ren:2013dzt}.
Different formulations of BChPT have been explored, including
the HB ChPT~\cite{Young:2009zb}, the EOMS BChPT~\cite{MartinCamalich:2010fp,Geng:2011wq,Ren:2012aj,Ren:2013dzt}, the
partial summation approach~\cite{Semke:2011ez,Semke:2012gs,Lutz:2012mq}, and the IR BChPT~\cite{Bruns:2012eh}.
In Refs.~\cite{MartinCamalich:2010fp,Geng:2011wq,Ren:2012aj,Ren:2013dzt}, we have performed a series of studies on the octet baryon masses by including finite-volume corrections (FVCs) self-consistently in the EOMS  BChPT up to next-to-next-to-leading order (NNLO) and N$^3$LO. In these studies, we found that the N$^3$LO EOMS BChPT can provide a good description of all the current LQCD data for the octet baryon masses, and we confirmed that the LQCD results are consistent with each other, although their setups are  different. Furthermore, the FVCs to the LQCD data are found to be important not only for the purpose of chiral extrapolations, but also for the determination of the corresponding LECs, especially for the many LECs appearing at N$^3$LO.

On the contrary,  there are only a few studies of the $n_f=2+1$
LQCD decuplet baryon masses~\cite{WalkerLoud:2008bp,Ishikawa:2009vc,Semke:2011ez,Semke:2012gs,Lutz:2012mq}. In Refs.~\cite{WalkerLoud:2008bp,Ishikawa:2009vc}, it was shown that the HB ChPT at NNLO cannot describe the LHPC and PACS-CS octet and
decuplet baryon masses. This  has motivated the series of studies on the $n_f=2+1$ LQCD octet baryon masses in
the EOMS framework~\cite{MartinCamalich:2010fp,Geng:2011wq,Ren:2012aj,Ren:2013dzt}.   The PACS-CS and LHPC decuplet baryon data
were also studied in Ref.~\cite{MartinCamalich:2010fp} up to NNLO  and a reasonable description of the LQCD data was achieved, contrary to the HB ChPT studies
of Refs.~\cite{WalkerLoud:2008bp,Ishikawa:2009vc}.  In Ref.~\cite{Semke:2011ez},
Semke and Lutz studied the BMW~\cite{Durr:2008zz} LQCD data  for the octet and decuplet baryon masses up to N$^3$LO in BChPT with the partial summation scheme.  It was shown that the light-quark mass dependence of the decuplet baryon masses can be well described. However, FVCs to the lattice data are not taken into account self-consistently. Whereas it has been shown in Refs.~\cite{Geng:2011wq,Ren:2012aj,Ren:2013dzt} that
FVCs need to be taken into account self-consistently in order to achieve a $\chi^2/\mathrm{d.o.f.}$ of about 1 in the description of the current $n_f=2+1$ LQCD octet baryon masses.

Given the fact that a simultaneous description of the $n_f=2+1$ LQCD decuplet baryon masses with FVCs taken into account self-consistently is still missing and that the EOMS BChPT can describe
the LQCD octet baryon masses rather well~\cite{MartinCamalich:2010fp,Geng:2011wq,Ren:2012aj,Ren:2013dzt}, it is timely to perform a thorough study of the lowest-lying decuplet baryon masses
in the EOMS BChPT up to N$^3$LO.
The paper is organized as follows:
In Sec.~\ref{SecII}, we collect the relevant chiral effective Lagrangians,
which contain $14$ to be determined LECs,
and calculate the decuplet baryon masses and the corresponding FVCs in covariant BChPT up to N$^3$LO.
In Sec.~\ref{SecIII}, we perform a simultaneous fit of the PACS-CS, QCDSF-UKQCD, and HSC data, study
the convergence of BChPT and the contributions of virtual octet and decuplet baryons, and compare the N$^3$LO BChPT with those LQCD data not included in the fit.
We then predict the pion- and strangeness-baryon sigma terms with the LECs determined from the best fits and compare them with the results of other
recent studies. A short summary is given in Sec.~\ref{SecIV}.

\section{Theoretical Framework}\label{SecII}

\subsection{Chiral effective Lagrangians}
The chiral effective Lagrangians relevant to the present study can be written as the sum of a mesonic part and a meson-baryon part:
\begin{equation}
  \mathcal{L}_{\rm eff} = \mathcal{L}_{\phi}^{(2)}+\mathcal{L}_{\phi}^{(4)} +\mathcal{L}_{\phi D}^{(1)}+\mathcal{L}_{\phi D}^{(2)}+\mathcal{L}_{\phi D}^{(4)}.
\end{equation}
The Lagrangians $\mathcal{L}_{\phi}^{(2)}$ and $\mathcal{L}_{\phi}^{(4)}$ of the mesonic sector can be found in Ref.~\cite{Ren:2012aj}. The leading order meson-baryon Lagrangian is
\begin{equation}
  \mathcal{L}_{\phi D}^{(1)} = \mathcal{L}_D + \mathcal{L}_{\phi BD}^{(1)} + \mathcal{L}_{\phi DD}^{(1)},
\end{equation}
where $\mathcal{L}_D$ denotes the covariant free Lagrangian, and $\mathcal{L}_{\phi BD}^{(1)}$ and $\mathcal{L}_{\phi DD}^{(1)}$ describe the interaction of the octet and decuplet baryons with the pseudoscalar mesons and have the following form:
\begin{eqnarray}
  \mathcal{L}_D&=&\bar{T}_{\mu}^{abc}\left(i\gamma^{\mu\nu\alpha}D_{\alpha} - m_D\gamma^{\mu\nu}\right)T_{\nu}^{abc},\label{Eq:freeLag}\\
  \mathcal{L}_{\phi BD}^{(1)}&=& \frac{i\mathcal{C}}{m_DF_{\phi}}\varepsilon^{abc}(\partial_{\alpha}\bar{T}_{\mu}^{ade}) \gamma^{\alpha\mu\nu}B_c^e\partial_{\nu}\phi_b^d + {\rm H.c.},\\
    \mathcal{L}_{\phi DD}^{(1)} &=& \frac{i\mathcal{H}}{m_DF_{\phi}}\bar{T}_{\mu}^{abc}\gamma^{\mu\nu\rho\sigma}\gamma_5 \left(\partial_{\rho}T_{\nu}^{abd}\right)\partial_{\sigma}\phi_{d}^c,
\end{eqnarray}
where we have used the so-called ``consistent'' coupling scheme for the meson-octet-decuplet vertices~\cite{Pascalutsa:1998pw,Pascalutsa:1999zz}. In the above Lagrangians, $m_D$ is the decuplet baryon mass in the chiral limit  and $T$ is the decuplet baryon field represented by the {\it Rarita-Schwinger} field $T^{abc}\equiv T_{\mu}^{abc}$. The physical fields are assigned as $T^{111}=\Delta^{++}$, $T^{112}=\Delta^+/\sqrt{3}$, $T^{122}=\Delta^0/\sqrt{3}$, $T^{222}=\Delta^-$,
$T^{113}=\Sigma^{*+}/\sqrt{3}$, $T^{123}=\Sigma^{*0}/\sqrt{6}$, $T^{223}=\Sigma^{*-}/\sqrt{3}$, $T^{133}=\Xi^{*0}/\sqrt{3}$, $T^{233}=\Xi^{*-}/\sqrt{3}$, and $T^{333}=\Omega^-$.  $D_{\nu}T_{\mu}^{abc}=\partial_{\nu}T_{\mu}^{abc} + (\Gamma_{\nu}, T_{\mu})^{abc}$, $ \Gamma_{\nu}=\frac{1}{2}\left\{u^{\dag} \partial_{\nu} u + u \partial_{\nu} u^{\dag}\right\}$ being the chiral connection with
$u={\rm exp}\left(i\frac{\phi}{2F_{\phi}}\right)$ collecting the pseudoscalar fields $\phi$,  and  $(X, T_{\mu})^{abc}\equiv (X)^a_dT_{\mu}^{dbc} + (X)^b_dT_{\mu}^{adc} + (X)^c_dT_{\mu}^{abd}$. The coefficient $F_{\phi}$ is the meson-decay constant in the chiral limit, and $\mathcal{C}$ ($\mathcal{H}$) denotes the $\phi BD$ ($\phi DD$) coupling. The totally antisymmetric gamma matrix products are defined as $ \gamma^{\mu\nu}=\frac{1}{2}[\gamma^\mu,\gamma^\nu]$,
$ \gamma^{\mu\nu\rho}=\frac{1}{2}\{\gamma^{\mu\nu},\gamma^{\rho}\}=-i\varepsilon^{\mu\nu\rho\sigma}\gamma_\sigma\gamma_5$,
$ \gamma^{\mu\nu\rho\sigma}=\frac{1}{2}[\gamma^{\mu\nu\rho},\gamma^\sigma]=i\varepsilon^{\mu\nu\rho\sigma}\gamma_5$
with the following conventions: $g^{\mu\nu}=\mathrm{diag}(1,-1,-1,-1)$,  $\varepsilon_{0,1,2,3}=-\varepsilon^{0,1,2,3}=1$,
$\gamma_5=i\gamma_0\gamma_1\gamma_2\gamma_3$~\cite{Geng:2008bm}. In the last and following Lagrangians, we sum over any repeated SU(3) index denoted by latin characters $a$, $b$, $c$, $\ldots$, and $(X)_b^a$ denotes the element of row $a$ and column $b$ of the matrix representation of $X$.

The meson-baryon Lagrangian at order $\mathcal{O}(p^2)$ can be written as
\begin{equation}
  \mathcal{L}_{\phi D}^{(2)} = \mathcal{L}_{\phi B}^{(2,~{\rm sb})} + \mathcal{L}_{\phi D}^{(2,~{\rm sb})} +
  {\mathcal{L}_{\phi D}^{(2)}}^{\prime}.
\end{equation}
The first and second terms denote the explicit chiral symmetry breaking part:
\begin{eqnarray}
    \mathcal{L}_{\phi B}^{(2,~{\rm sb})}&=&b_0\langle\bar{B}B\rangle\langle\chi_+\rangle + b_{D/F}\langle\bar{B}\left[\chi_{+},B\right]_{\pm}\rangle,\\
    \mathcal{L}_{\phi D}^{(2,~{\rm sb})}&=&\frac{t_0}{2}\bar{T}_{\mu}^{abc}g^{\mu\nu}T_{\nu}^{abc}\langle\chi_{+}\rangle + \frac{t_D}{2}\bar{T}_{\mu}^{abc}g^{\mu\nu}(\chi_+, T_{\nu})^{abc},
\end{eqnarray}
where $b_0$, $b_D$, $b_F$, $t_0$, and $t_D$ are the LECs, $\chi_+=u^{\dag}\chi u^{\dag}+ u\chi^{\dag}u$ , and $\chi=2 B_0\mathcal{M}$ accounts for explicit chiral symmetry breaking
with $B_0=-\langle0|\bar{q}q|0\rangle/F_{\phi}^2$ and $\mathcal{M}={\rm diag}(m_l,~m_l,~m_s)$.
 For the chiral symmetry conserving part $ {\mathcal{L}_{\phi D}^{(2)}}^{\prime}$,
 one has nine terms, following the conventions of  Refs.~\cite{Lutz:2010se,Semke:2011ez},
\begin{eqnarray}
{\mathcal{L}_{\phi D}^{(2)}}^{\prime}&=& \frac{1}{F_{\phi}^2}\left\{ t_1 \bar{T}_{\mu}^{abc}g^{\mu\nu}
  \left(\partial^{\sigma}\phi\partial_{\sigma}\phi\right)_c^dT_{\nu}^{abd} +
  t_2 \bar{T}_{\mu}^{abc}
  \left[\left(\partial^{\mu}\phi\partial_{\nu}\phi\right)_c^d+
  \left(\partial_{\nu}\phi\partial^{\mu}\phi\right)_c^d\right]T^{\nu,abd}\right.\nonumber\\
  && + t_3 \bar{T}_{\mu,abc}\partial_{\nu}\phi_d^a\varepsilon^{bde} T^{\mu,fgc}\partial^{\nu}\phi_f^h\varepsilon_{ghe} + t_4\bar{T}_{\mu,abc}\left[\partial^{\mu}\phi_d^a\varepsilon^{bde} \partial_{\nu}\phi_f^h\varepsilon_{ghe} + \partial_{\nu}\phi_d^a\varepsilon^{bde} \partial^{\mu}\phi_f^h\varepsilon_{ghe}\right]T^{\nu,fgc}\nonumber\\
  && + t_5 \bar{T}_{\mu}^{abc}g^{\mu\nu}T_{\nu}^{abc}\langle\partial^{\sigma}\phi\partial_{\sigma}\phi\rangle + t_6 \bar{T}_{\mu}^{abc}T^{\nu,abc}\langle\partial^{\mu}\phi\partial_{\nu}\phi\rangle \nonumber\\
  && + t_7\left[\left(\bar{T}_{\alpha}^{abc}\left(\partial_{\mu}\phi\partial_{\nu}\phi\right)_c^d i\gamma^{\mu}\partial^{\nu}T^{\alpha,abd} + \bar{T}_{\alpha}^{abc}\left(\partial_{\nu}\phi\partial_{\mu}\phi\right)_c^d i\gamma^{\mu}\partial^{\nu}T^{\alpha,abd}\right) + {\rm H.c.} \right]\nonumber\\
  && + t_8\left[\left(\bar{T}_{\alpha,abc}\partial_{\mu}\phi_d^a\varepsilon^{bde}i\gamma^{\mu} \partial^{\nu}T^{\alpha,fgc}\partial_{\nu}\phi_f^h\varepsilon_{ghe} + \bar{T}_{\alpha,abc}\partial_{\nu}\phi_d^a\varepsilon^{bde}i\gamma^{\mu} \partial^{\nu}T^{\alpha,fgc}\partial_{\mu}\phi_f^h\varepsilon_{ghe} \right) + {\rm H.c.} \right]\nonumber\\
  &&\left. +  t_9\left[\bar{T}_{\alpha}^{abc}i\gamma^{\mu}\partial^{\nu}T^{\alpha,abc}
  \langle\partial_{\mu}\phi\partial_{\nu}\phi\rangle + {\rm H.c.}\right] \right\},
\end{eqnarray}
where $t_{1,\dots,6}$ have dimension mass$^{-1}$ and $t_{7,\dots,9}$ have dimension mass$^{-2}$.

The fourth order chiral effective Lagrangians contain five LECs  (see also  Refs.~\cite{Tiburzi:2004rh,Semke:2011ez}):
\begin{eqnarray}
  \mathcal{L}_{\phi D}^{(4)} &= & e_1\bar{T}_{\mu}^{abc}g^{\mu\nu}\left(\chi_+^2\right)_d^cT_{\nu}^{abd} + e_2\left(\bar{T}_{\mu}^{abc}\left(\chi_+\right)_c^d\right)g^{\mu\nu} \left(\left(\chi_+\right)_e^bT_{\nu}^{aed}\right)\nonumber\\
  && + e_3\bar{T}_{\mu}^{abc}g^{\mu\nu}\left(\chi_+\right)_d^cT_{\nu}^{abd}\langle\chi_+\rangle + e_4\bar{T}_{\mu}^{abc}g^{\mu\nu}T_{\nu}^{abc}\langle\chi_+\rangle^2\nonumber\\
  && + e_5\bar{T}_{\mu}^{abc}g^{\mu\nu}T_{\nu}^{abc}\langle\chi_+^2\rangle.
\end{eqnarray}

The propagator of the spin-$3/2$ fields in $d$ dimensions has the following form~\cite{Pascalutsa:2005nd}:
\begin{equation}
  S^{\mu\nu}(p) = -\frac{\slashed p+ m_D}{p^2-m_D^2+i\epsilon} \left[g^{\mu\nu}-\frac{1}{d-1}\gamma^{\mu}\gamma^{\nu} - \frac{1}{(d-1)m_D}\left(\gamma^{\mu}p^{\nu}-\gamma^{\nu}p^{\mu}\right) - \frac{d-2}{(d-1)m_D^2}p^{\mu}p^{\nu}\right].
\end{equation}

\begin{figure}[t]
  \centering
  \includegraphics[width=16cm]{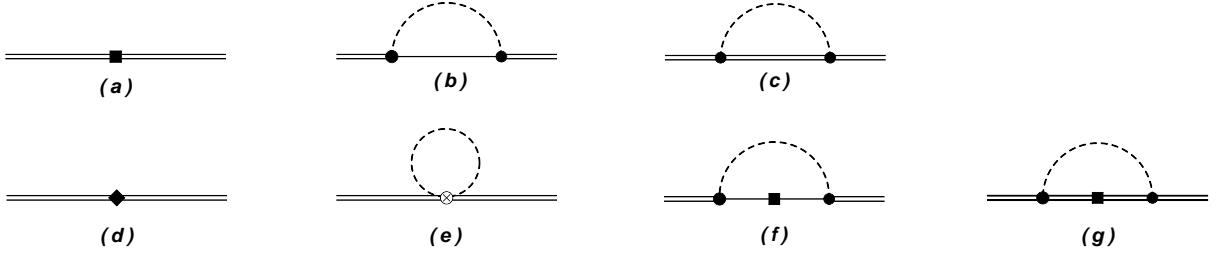}\\
  \caption{Feynman diagrams contributing to the decuplet baryon masses up to $\mathcal{O}(p^4)$ in the EOMS BChPT. The single lines correspond to octet baryons, double lines to decuplet baryons and dashed lines to mesons. The black boxes (diamond) indicate second (fourth) order couplings. The solid dot (circle-cross) indicates an insertion from the dimension one (two) meson-baryon Lagrangians. Wave function renormalization diagrams are not explicitly shown but included in the calculation.}
  \label{Fig:FeyDia}
\end{figure}

\subsection{Decuplet baryon masses}
In this subsection, the decuplet baryon masses are calculated in the limit of exact isospin symmetry. Formally, up to $\mathcal{O}(p^4)$ the baryon masses can be written as
\begin{equation}\label{eq:decmass}
  M_D = m_{D} + m_D^{(2)} + m_D^{(3)} + m_D^{(4)},
\end{equation}
where $m_D$ is the decuplet baryon mass in the chiral limit. The $m_D^{(2)}$, $m_D^{(3)}$, and $m_D^{(4)}$ are  the next-to-leading order (NLO),  NNLO, and N$^3$LO chiral corrections to the decuplet baryon masses, respectively. The corresponding Feynman diagrams are shown in Fig.~\ref{Fig:FeyDia}, and the explicit expression of the decuplet baryon masses is
\begin{eqnarray}\label{Eq:decmass2}
  M_D &=& m_D + \xi_{D,\pi}^{(a)}M_{\pi}^2 + \xi_{D,K}^{(a)}M_{K}^2\nonumber\\
      &&+ \frac{1}{(4\pi F_{\phi})^2}\sum\limits_{\phi=\pi,~K,~\eta} \left[ \xi_{D,\phi}^{(b)}H_D^{(b)}(M_{\phi}) +\xi_{D,\phi}^{(c)}H_D^{(c)}(M_{\phi})\right] \nonumber\\
      &&+\xi_{D,\pi}^{(d)}M_{\pi}^4 + \xi_{D,K}^{(d)}M_K^4 + \xi_{D,\pi K}^{(d)}M_{\pi}^2M_{K}^2\nonumber\\
      &&+ \frac{1}{(4\pi F_{\phi})^2}\sum\limits_{\phi=\pi,~K,~\eta}
        \left[\xi_{D,\phi}^{(e,1)}H_D^{(e,1)}(M_{\phi})+\xi_{D,\phi}^{(e,2)}H_D^{(e,2)}(M_{\phi}) +\xi_{D,\phi}^{(e,3)}H_D^{(e,3)}(M_{\phi})\right]\nonumber\\
      && - \frac{1}{(4\pi F_{\phi})^2}\sum_{\substack{\phi=\pi,~K,~\eta\\  B=N,~\Lambda,~\Sigma,~\Xi}}\xi_{DB,\phi}^{(f)} H_{D,B}^{(f)}(M_{\phi}) \nonumber\\
      && - \frac{1}{(4\pi F_{\phi})^2} \sum_{\substack{\phi=\pi,~K,~\eta\\D^{\prime}=\Delta,~\Sigma^*,~\Xi^*,~\Omega^-}} \xi_{DD^{\prime},\phi}^{(g)} H_{D,D^{\prime}}^{(g)}(M_{\phi}),
\end{eqnarray}
where $\xi^{(i)}$'s and $H^{(i)}$'s are the corresponding coefficients and loop functions with the subscript $i$ denoting the corresponding
diagrams shown in Fig.~\ref{Fig:FeyDia}. The $\xi^{(i)}$'s are tabulated in Tables \ref{Tab:NLO} and \ref{Tab:N3LO}.

\begin{table}[t]
  \centering
  \caption{Coefficients of the NLO and NNLO contributions to the decuplet baryon masses [Eq.~\eqref{Eq:decmass2}].}
  \label{Tab:NLO}
  \begin{tabular}{ccccc}
    \hline\hline
      & $\Delta$ & $\Sigma^*$ & $\Xi^*$ & $\Omega^-$ \\
    \hline
    $\xi_{D,\pi}^{(a)}$ & $t_0+3t_D$ & $t_0+t_D$ & $t_0-t_D$ & $t_0-3t_D$ \\
    $\xi_{D,K}^{(a)}$ & $2t_0$ & $2t_0+2t_D$ & $2t_0+4t_D$ & $2t_0+6t_D$ \\
    \hline
    $\xi_{D,\pi}^{(b)}$ & $\frac{4}{3}\mathcal{C}^2$ & $\frac{10}{9}\mathcal{C}^2$ & $\frac{2}{3}\mathcal{C}^2$ & $0$ \\
    $\xi_{D,K}^{(b)}$ & $\frac{4}{3}\mathcal{C}^2$ & $\frac{8}{9}\mathcal{C}^2$ & $\frac{4}{3}\mathcal{C}^2$ & $\frac{8}{3}\mathcal{C}^2$ \\
    $\xi_{D,\eta}^{(b)}$ & $0$ & $\frac{2}{3}\mathcal{C}^2$ & $\frac{2}{3}\mathcal{C}^2$ & $0$ \\
    \hline
    $\xi_{D,\pi}^{(c)}$ & $\frac{50}{27}\mathcal{H}^2$ & $\frac{80}{81}\mathcal{H}^2$ & $\frac{10}{27}\mathcal{H}^2$ & $0$ \\
    $\xi_{D,K}^{(c)}$ & $\frac{20}{27}\mathcal{H}^2$ & $\frac{160}{81}\mathcal{H}^2$ & $\frac{20}{9}\mathcal{H}^2$ & $\frac{40}{27}\mathcal{H}^2$ \\
    $\xi_{D,\eta}^{(c)}$ & $\frac{10}{27}\mathcal{H}^2$ & $0$ & $\frac{10}{27}\mathcal{H}^2$ & $\frac{40}{27}\mathcal{H}^2$ \\
    \hline\hline
  \end{tabular}
\end{table}

\begin{table}[t]
\scriptsize
  \centering
  \caption{Coefficients of the  N$^3$LO contributions to the decuplet baryon masses  [Eq.~\eqref{Eq:decmass2}], with
  the following notations: $\tilde{t}_1=2t_1+t_2$, $\tilde{t}_2=2t_3+t_4$, and $\tilde{t}_3=4t_5+t_6$.}
  \label{Tab:N3LO}
  \begin{tabular}{ccccc}
    \hline\hline
      & $\Delta$ & $\Sigma^*$ & $\Xi^*$ & $\Omega^-$ \\
    \hline
    $\xi_{D,\pi}^{(d)}$ & $4(e_1+e_2+e_3+e_4+3e_5)$ & $\frac{4}{3}(3e_1-e_2+e_3+3e_4+9e_5)$ & $\frac{4}{3}(3e_1-e_2-e_3+3e_4+9e_5)$ & $4(e_1+e_2-e_3+e_4+3e_5)$ \\
    $\xi_{D,K}^{(d)}$ & $16(e_4+e_5)$ & $\frac{16}{3}(e_1+e_3+3e_4+3e_5)$ & $\frac{16}{3}(2e_1+e_2+2e_3+3e_4+3e_5)$ & $16(e_1+e_2+e_3+e_4+e_5)$ \\
    $\xi_{D,\pi K}^{(d)}$ & $8(e_3+2e_4-2e_5)$ & $-\frac{16}{3}(e_1-e_2-e_3-3e_4+3e_5)$ & $-\frac{8}{3}(4e_1-e_3-6e_4+6e_5)$ & $-16(e_1+e_2-e_4+e_5)$ \\
    \hline
    $\xi_{D,\pi}^{(e,1)}$  & $\frac{3}{2}(2t_0+3t_D)M_{\pi}^2$ & $3(t_0+t_D)M_{\pi}^2$ & $\frac{3}{2}(2t_0+t_D)M_{\pi}^2$ & $3t_0M_{\pi}^2$ \\
    $\xi_{D,K}^{(e,1)}$    & $(4t_0+3t_D)M_{K}^2$ & $4(t_0+t_D)M_{K}^2$ & $(4t_0+5t_D)M_{K}^2$ & $2(2t_0+3t_D)M_{K}^2$ \\
    $\xi_{D,\eta}^{(e,1)}$ & $\frac{1}{6}\left[8t_0M_{K}^2-(2t_0-3t_D)M_{\pi}^2\right]$ & $\frac{1}{3}(t_0+t_D)(4M_{K}^2-M_{\pi}^2)$ & $\frac{1}{6}\left[8(t_0+2t_D)M_{K}^2-(2t_0+7t_D)M_{\pi}^2\right]$ & $\frac{1}{3}\left[4(t_0+3t_D)M_{K}^2-(t_0+6t_D)M_{\pi}^2\right]$ \\
    \hline
    $\xi_{D,\pi}^{(e,2)}$ & $-\frac{1}{2}(3\tilde{t}_1+2\tilde{t}_2+3\tilde{t}_3)$ & $-\frac{1}{6}(6\tilde{t}_1+5\tilde{t}_2+9\tilde{t}_3)$ & $-\frac{1}{2}(\tilde{t}_1+\tilde{t}_2+3\tilde{t}_3)$ & $-\frac{3}{2}\tilde{t}_3$\\
    $\xi_{D,K}^{(e,2)}$ & $-(\tilde{t}_1+\tilde{t}_2+2\tilde{t}_3)$ & $-\frac{2}{3}(2\tilde{t}_1+\tilde{t}_2+3\tilde{t}_3)$ & $-\frac{1}{3}(5\tilde{t}_1+3\tilde{t}_2+6\tilde{t}_3)$ & $-2(\tilde{t}_1+\tilde{t}_2+\tilde{t}_3)$\\
    $\xi_{D,\eta}^{(e,2)}$ & $-\frac{1}{6}(\tilde{t}_1+3\tilde{t}_3)$ & $-\frac{1}{6}(2\tilde{t}_1+3\tilde{t}_2+3\tilde{t}_3)$ & $-\frac{1}{2}(\tilde{t}_1+\tilde{t}_2+\tilde{t}_3)$& $-\frac{1}{6}(4\tilde{t}_1+3\tilde{t}_3)$ \\
    \hline
    $\xi_{D,\pi}^{(e,3)}$ & $-4(3t_7+2t_8+3t_9)$ & $-\frac{4}{3}(6t_7+5t_8+9t_9)$ & $-4(t_7+t_8+3t_9)$ & $-12t_9$\\
    $\xi_{D,K}^{(e,3)}$ & $-8(t_7+t_8+2t_9)$ & $-\frac{16}{3}(2t_7+t_8+3t_9)$ & $-\frac{8}{3}(5t_7+3t_8+6t_9)$ & $-16(t_7+t_8+t_9)$\\
    $\xi_{D,\eta}^{(e,3)}$ & $-\frac{4}{3}(t_7+3t_9)$ & $-\frac{4}{3}(2t_7+3t_8+3t_9)$ & $-4(t_7+t_8+t_9)$& $-\frac{4}{3}(4t_7+3t_9)$ \\
    \hline
    $\xi_{D N,\{\pi,K,\eta\}}^{(f)}$ & $\{2\mathcal{C}^2,0,0\}$ & $\{0,\frac{2}{3}\mathcal{C}^2,0\}$ & $\{0,0,0\}$ & $\{0,0,0\}$\\
    $\xi_{D \Lambda,\{\pi,K,\eta\}}^{(f)}$ & $\{0,0,0\}$ & $\{\mathcal{C}^2,0,0\}$ & $\{0,\mathcal{C}^2,0\}$ & $\{0,0,0\}$\\
    $\xi_{D \Sigma,\{\pi,K,\eta\}}^{(f)}$ & $\{0,2\mathcal{C}^2,0\}$ & $\{\frac{2}{3}\mathcal{C}^2,0,\mathcal{C}^2\}$ & $\{0,\mathcal{C}^2,0\}$ & $\{0,0,0\}$\\
    $\xi_{D \Xi,\{\pi,K,\eta\}}^{(f)}$ & $\{0,0,0\}$ & $\{0,\frac{2}{3}\mathcal{C}^2,0\}$ & $\{\mathcal{C}^2,0,\mathcal{C}^2\}$ & $\{0,0,4\mathcal{C}^2\}$\\
    \hline
    $\xi_{D \Delta,\{\pi,K,\eta\}}^{(g)}$ & $\{\frac{5}{3}\mathcal{H}^2,0,\frac{1}{3}\mathcal{H}^2\}$ & $\{0,\frac{8}{9}\mathcal{H}^2,0\}$ & $\{0,0,0\}$ & $\{0,0,0\}$\\
    $\xi_{D \Sigma^*,\{\pi,K,\eta\}}^{(g)}$ & $\{0,\frac{2}{3}\mathcal{H}^2,0\}$ & $\{\frac{8}{9}\mathcal{H}^2,0,0\}$ & $\{0,\frac{4}{3}\mathcal{H}^2,0\}$ & $\{0,0,0\}$\\
    $\xi_{D \Xi^*,\{\pi,K,\eta\}}^{(g)}$ & $\{0,0,0\}$ & $\{0,\frac{8}{9}\mathcal{H}^2,0\}$ & $\{\frac{1}{3}\mathcal{H}^2,0,\frac{1}{3}\mathcal{H}^2\}$ & $\{0,\frac{4}{3}\mathcal{H}^2,0\}$\\
    $\xi_{D \Omega^-,\{\pi,K,\eta\}}^{(g)}$ & $\{0,0,0\}$ & $\{0,0,0\}$ & $\{0,\frac{2}{3}\mathcal{H}^2,0\}$ & $\{0,0,\frac{4}{3}\mathcal{H}^2\}$\\
    \hline\hline
  \end{tabular}
\end{table}

In Eq.~\eqref{Eq:decmass2}, the loop functions $H_{D}^{(b)}$, $H_{D}^{(c)}$, $H_D^{(e,1)}$, $H_D^{(e,2)}$, $H_D^{(e,3)}$, $H_{D,B}^{(f)}$, and $H_{D,D^\prime}^{(g)}$ are obtained by using the $\overline{\rm MS}$ renormalization scheme to remove the divergent pieces and the EOMS renormalization scheme to remove the power-counting-breaking (PCB) terms~\cite{Gegelia:1999gf,Fuchs:2003qc,Geng:2013xn}. The explicit expressions of $H_{D}^{(b)}$, $H_{D}^{(c)}$ can be found in Ref.~\cite{MartinCamalich:2010fp}, and the others are given in
the Appendix.

It should be noted that in the evaluation of the diagrams in Figs.~\ref{Fig:FeyDia}(f) and (g), we have only kept terms linear in $M_D^{(2)}$ and $M_B^{(2)}$, in accordance with
our power counting.
At N$^3$LO, the pseudoscalar meson masses appearing in $m_D^{(2)}$ should be replaced by their $\mathcal{O}(p^4)$ counterparts to generate the N$^3$LO contributions to $m_D^{(4)}$.
The explicit expressions of the meson masses up to $\mathcal{O}(p^4)$ can be found in Ref.~\cite{Gasser:1984gg}. The empirical values of the LECs $L_i^{r}$ $(i=1,\dots,10)$ are taken from the latest global fit~\cite{Bijnens:2011tb}. In order to be consistent with our renormalization scale used for the baryon sector, we have reevaluated the LECs at $\mu=1$ GeV. The details can be found in Ref.~\cite{Ren:2012aj}.~\footnote{
 In both Ref.~\cite{Ren:2012aj} and the present work, the LQCD pseudoscalar masses are treated as LO masses. We have checked that treating them as NLO masses does not
 affect in any significant way the results of both studies.}

\subsection{Finite-volume corrections}
As emphasized in Refs.~\cite{Geng:2011wq,Ren:2012aj,Ren:2013dzt},  FVCs have
to be taken into account in studying the current LQCD data.  In the case of the decuplet baryon masses, they have been studied up to NNLO in
the EOMS BChPT~\cite{MartinCamalich:2010fp} and in the HB ChPT~\cite{Ishikawa:2009vc}. In the following, we extend the study up to N$^3$LO in
the EOMS  BChPT.

The FVCs can be easily evaluated following  the standard technique. One chooses the baryon rest frame, i.e., $p^{\mu}=(m_D,\vec{0})$, performs a momentum shift and wick rotation,
integrates over the temporal dimension, and obtains the results expressed in terms of the master formulas given in Ref.~\cite{Beane:2004tw}. See Refs.\cite{AliKhan:2003cu,Beane:2004tw,Geng:2011wq} for more details.

To proceed with the above procedure, one should note that since
Lorentz invariance is lost in finite volume, the mass term in the loop functions is identified as the term having the structure of
$\delta_{ij}$. This can be easily seen by noticing that at the rest frame the zero component of the decuplet baryon field vanishes because of the
on-shell condition $p_\mu T^\mu=0$.  For instance,  the loop function of  the diagram  in Fig.~\ref{Fig:FeyDia}(b), after Feynman parametrization, becomes
\begin{equation}\label{Eq:intFVCB}
  G^{(b)}_D \propto \int\frac{d^4 k}{(2\pi)^4} \frac{(m_D(x-1)-m_B)k^{\alpha}k^{\beta}}{\left(k^2-{\mathcal{M}_D^{(b)}}^2\right)^2},
\end{equation}
where ${\mathcal{M}_D^{(b)}}^2 = (x^2-x)m_D^2 + xm_0^2 + (1-x)M_{\phi}^2 - i\epsilon$. To evaluate its contribution to the decuplet baryon mass, one simply replaces $k^{\alpha}k^{\beta}$ with $\delta_{ij}\vec{k}^2/3$ in the numerator.  Following the procedure specified
above, one can then easily obtain the FVCs to the loop function of the diagram in Fig.~\ref{Fig:FeyDia}(b),
\begin{eqnarray}\label{Eq:FVCsB}
  \delta G_{D}^{(b)}(M_{\phi})& \equiv& G_D^{(b)}(L)- G_D^{(b)}(\infty)\nonumber\\
  &=& -\frac{1}{12} \int_0^1dx~\left[m_0-m_D(x-1)\right] \left[\delta_{1/2}({\mathcal{M}_D^{(b)}}^2)- {\mathcal{M}_D^{(b)}}^2\delta_{3/2}({\mathcal{M}_D^{(b)}}^2)\right],
\end{eqnarray}
where the ``master" formulas $\delta_r (\mathcal{M}^2)$ are defined as
\begin{equation}\label{eq:master}
\delta_r (\mathcal{M}^2)=
\frac{2^{-1/2-r}(\sqrt{\mathcal{M}^2})^{3-2r}}{\pi^{3/2}\Gamma(r)}
\sum_{\vec{n}\ne0}(L\sqrt{\mathcal{M}^2}|\vec{n}|)^{-3/2+r}K_{3/2-r}(L\sqrt{\mathcal{M}^2}|\vec{n}|),
\end{equation}
where $K_n(z)$ is the modified Bessel function of the second kind, and $\sum\limits_{\vec{n}\ne0}\equiv\sum\limits^{\infty}_{n_x=-\infty}\sum\limits^{\infty}_{n_y=-\infty}\sum\limits^{\infty}_{n_z=-\infty}(1-\delta(|\vec{n}|,0))$ with $\vec{n}=(n_x,n_y,n_z)$.

Following the same procedure, one can obtain the FVCs of the other loop diagrams in Fig.~\ref{Fig:FeyDia}. For the NNLO one-loop diagram of Fig.~\ref{Fig:FeyDia}(c), one obtains
\begin{equation}\label{Eq:FVCsC}
  \delta G_{D}^{(c)}(M_{\phi}) = \frac{5}{36} \int_0^1dx~m_D(x-2)\left[\delta_{1/2}({\mathcal{M}_D^{(c)}}^2) - {\mathcal{M}_D^{(c)}}^2\delta_{3/2}({\mathcal{M}_D^{(c)}}^2)\right],
\end{equation}
with ${\mathcal{M}_D^{(c)}}^2 = x^2m_D^2 + (1-x)M_{\phi}^2 - i\epsilon.$
Taking the limit of  $m_D\rightarrow\infty$, Eq.~\eqref{Eq:FVCsB} and Eq.~\eqref{Eq:FVCsC} reduce to
\begin{eqnarray}
  \delta G_{D}^{(b)}(M_{\phi})_{\rm HB} &=& -\frac{1}{8}\int_0^{\infty} dx \left[\delta_{1/2}(\beta_{\Delta}^2)-\beta_{\Delta}^2\delta_{3/2}(\beta_{\Delta}^2)\right],\\
  \delta G_{D}^{(c)}(M_{\phi})_{\rm HB} &=& \frac{1}{2}\int_0^{\infty} dx
  \left[\delta_{1/2}(\beta^2)-\beta^2\delta_{3/2}(\beta^2)\right],
\end{eqnarray}
where $\beta_{\Delta}=x^2-2x\delta+M_{\phi}^2$ and $\beta=x^2+M_{\phi}^2$.  They agree with the HB ChPT results of Ref.~\cite{Ishikawa:2009vc}.

FVCs to the N$^3$LO one-loop diagrams in Figs.~\ref{Fig:FeyDia}~(e), (f), and (g) have the following form:
\begin{eqnarray}
  \delta G_{D}^{(e,1)}(M_{\phi}) &=& \frac{1}{2}\delta_{1/2}(M_{\phi}^2),\\
  \delta G_{D}^{(e,2)}(M_{\phi}) &=& \frac{1}{2}M_{\phi}^2\delta_{1/2}(M_{\phi}^2),\\
  \delta G_{D}^{(e,3)}(M_{\phi}) &=& \frac{1}{2}m_D\delta_{-1/2}(M_{\phi}^2),
\end{eqnarray}
\begin{eqnarray}\label{Eq:FVCsF}
  \delta G_{D,B}^{(f)}(M_{\phi}) &=& \frac{1}{12}\int_0^1 dx \left\{\left[m_D^{(2)}(x-1)-m_B^{(2)}\right]\cdot \delta_{1/2}({\mathcal{M}_D^{(b)}}^2)\right.\nonumber\\
  && +\left[(1-x)\left(2{\mathcal{M}_D^{(b)}}^2+M_{\phi}^2(x-1)- (m_{0}+m_{D})(m_0+2m_D)x+2m_D^2x^2\right)m_D^{(2)}\right.\nonumber\\
  && \quad + \left.\left({\mathcal{M}_D^{(b)}}^2+3m_0x(m_0+m_D(1-x))\right)m_B^{(2)}\right]\cdot \delta_{3/2}({\mathcal{M}_D^{(b)}}^2)\nonumber\\
  && +{\mathcal{M}_D^{(b)}}^2\left[(x-1)\left({\mathcal{M}_D^{(b)}}^2 + M_{\phi}^2(x-1)-(m_0+m_D)(m_0+2m_D)x+2m_D^2x^2\right)m_D^{(2)}\right.\nonumber\\
  &&\left.\left.\quad - 3m_0x(m_0+m_D(1-x))m_B^{(2)}\right]\cdot\delta_{5/2}({\mathcal{M}_D^{(b)}}^2)\right\},
\end{eqnarray}
\begin{eqnarray}\label{Eq:decmassG}
  \delta G_{D,D^{\prime}}^{(g)}(M_{\phi}) &=& \frac{5}{36}\int_0^1 dx
  \left\{\left[(3x-5)m_D^{(2)} + (3-2x)m_{D^{\prime}}^{(2)}\right] \cdot\delta_{1/2}({\mathcal{M}_{D}^{(c)}}^2)\right.\nonumber\\
  && +\left[\left({\mathcal{M}_D^{(c)}}^2(10-6x)+2m_D^2x(2x-3)+M_{\phi}^2(-3x^2+8x-5)\right) m_{D}^{(2)}\right.\nonumber\\
  && \quad + \left.\left({\mathcal{M}_D^{(c)}}^2(2x-3)-3m_D^2(x^2-2x)\right)m_{D^{\prime}}^{(2)}\right]
  \cdot\delta_{3/2}({\mathcal{M}_{D}^{(c)}}^2)\nonumber\\
  && + \left[{\mathcal{M}_D^{(c)}}^2\left(m_D^2(6x-4x^2)+{\mathcal{M}_D^{(c)}}^2(3x-5)+M_{\phi}^2(x-1) (3x-5)\right)m_{D}^{(2)}\right. \nonumber\\
  &&\quad + \left.\left.3{\mathcal{M}_D^{(c)}}^2m_D^2(x^2-2x)m_{D^{\prime}}^{(2)}\right]
  \cdot\delta_{5/2}({\mathcal{M}_{D}^{(c)}}^2)\right\}.
\end{eqnarray}

The above standard procedure applies only to the case where  $m_D\le m_0+M_\phi$. For the case of $m_D>m_0+M_\phi$, we follow the approach proposed in Ref.~\cite{Bernard:2007cm} and replace
the original   $\delta_r(\mathcal{M}^2)$ with three parts by introducing a new scale $\mu$ satisfying $\mu<m_0+M_{\phi}$, i.e.,
\begin{equation}
  \delta_r(\mathcal{M}^2)=g_1^r - g_2^r + g_3^r,
\end{equation}
where the $g_{1,2,3}^r$ are defined as
\begin{eqnarray}
  g_1^r &=& \frac{1}{L^3}\sum\limits_{\vec{k}}\left\{\frac{1}{\left[\frac{4\pi^2\vec{n}^2}{L^2}+ \mathcal{M}^2(m_D^2)\right]^r} -\frac{1}{\left[\frac{4\pi^2\vec{n}^2}{L^2}+\mathcal{M}^2(\mu^2)\right]^r} + \frac{r(x^2-x)(m_D^2-\mu^2)}{\left[\frac{4\pi^2\vec{n}^2}{L^2}+\mathcal{M}^2(\mu^2)\right]^{r+1}}\right\},\\
  g_2^r &=& \int_0^{+\infty}\frac{k^2dk}{2\pi^2}\cdot \left\{\frac{1}{\left[\vec{k}^2+\mathcal{M}^2(m_D^2)\right]^r} -\frac{1}{\left[\vec{k}^2+\mathcal{M}^2(\mu^2)\right]^r}+ \frac{r(x^2-x)(m_D^2-\mu^2)}{\left[
  \vec{k}^2+\mathcal{M}^2(\mu^2)\right]^{r+1}}\right\},\\
  g_3^r &=& \delta_{r}\left(\mathcal{M}^2(\mu^2)\right) - r(x^2-x)(m_D^2-\mu^2)\delta_{r+1}\left(\mathcal{M}^2(\mu^2)\right),
\end{eqnarray}
with
\begin{eqnarray}
  \mathcal{M}^2(m_D^2) &=& (x^2-x)m_D^2 + xm_0^2 + (1-x)M_{\phi}^2 - i\epsilon,\\
\mathcal{M}^2(\mu^2) &=& (x^2-x)\mu^2 + xm_0^2 + (1-x)M_{\phi}^2 - i\epsilon.
\end{eqnarray}

To take into account the FVCs in the study of the LQCD data,
one simply replaces the loop functions $H$ of Eq.~\eqref{Eq:decmass2} by $\tilde{H}=H+\delta G$ with the $\delta G$s calculated above.

\section{Results and discussion}\label{SecIII}
In this section, we perform a simultaneous fit of the $n_f=2+1$ LQCD data from the PACS-CS~\cite{Aoki:2008sm}, QCDSF-UKQCD~\cite{Bietenholz:2011qq}, and HSC~\cite{Lin:2008pr} Collaborations and
the experimental data~\cite{Beringer:1900zz} to determine the $17$ unknown LECs, $m_D$, $t_D$, $t_{0\cdots9}$, and $e_{1\cdots5}$.  Since
$t_1$, $t_2$, $t_3$, $t_4$, $t_5$, and $t_6$ appear in combinations, effectively we have only 14 independent LECs. The pion or light-quark mass dependence of the decuplet baryon masses is studied in the NLO, NNLO, and N$^3$LO EOMS BChPT. Using the so-obtained LECs, we also carry out a detailed study on the QCDSF-UKQCD and LHPC data to test the applicability of the N$^3$LO BChPT and the consistency between different LQCD simulations. Furthermore, the pion- and strangeness-baryon sigma terms are  predicted by the use of the Feynman-Hellmann theorem.

\subsection{LQCD data and values of LECs}

Up to now, five collaborations have reported $n_f=2+1$ simulations of the decuplet baryon masses, i.e., the BMW~\cite{Durr:2008zz}, PACS-CS~\cite{Aoki:2008sm}, LHPC~\cite{WalkerLoud:2008bp}, HSC~\cite{Lin:2008pr}, and  QCDSF-UKQCD~\cite{Bietenholz:2011qq} Collaborations. Because the BMW data are not publicly available and the
 data of the LHPC Collaboration seem to suffer some systematic errors, as shown in their chiral extrapolation result on the $\Delta$(1232) mass, which is much higher than its physical value~\cite{WalkerLoud:2008bp} (see also Sec. \ref{subsec3b}),
 we will concentrate on the data of the
 PACS-CS,  QCDSF-UKQCD, and HSC  Collaborations. Following the criteria used in our previous studies~\cite{Ren:2013dzt}, we only select the LQCD data that satisfy  $M_\pi<0.5$ GeV and $M_{\phi}L>3.8$. As a result, there are eight sets of data from the PACS-CS (3 sets), QCDSF-UKQCD (2 sets), and HSC (3 sets) Collaborations. Among the eight LQCD data sets studied, only in the ensemble with $M_{\pi}=296$ MeV  from the PACS-CS Collaboration, can the decay $\Delta\rightarrow N+\pi$ happen. It should be noted
 that the PACS-CS Collaboration measured the lowest energy levels of the vector meson and decuplet baryon channels, which are different from the true resonance masses. The resulting difference for the $\rho$ meson is estimated to be 5 percent using L\"{u}scher's formula~\cite{Aoki:2008sm}. We will comment on this later.

It should be mentioned that the $\mathcal{O}(a)$-improved Wilson action was used by all the above collaborations except the LHPC Collaboration, which employed a mixed action. The $\mathcal{O}(a)$-improved action has the favorable property that the leading order corrections from the finite lattice spacing are eliminated. The finite lattice spacing corrections of the mixed action of the LHPC Collaboration were also shown to be small~\cite{WalkerLoud:2008bp}. Therefore, in the present work we assume that the discretization artifacts of the present LQCD simulations are small  and can be ignored, and will leave a detailed study on finite lattice spacing artifacts to a future study (for a recent study of the discretization effects on the octet baryon masses, see Ref.~\cite{Ren:2013wxa}).

Before we perform a simultaneous fit of the LQCD data, we specify our strategy to fix some of the LECs in the N$^3$LO BChPT mass formulas [Eq.~\eqref{Eq:decmass2}]. For the meson-decay constant, we use $F_{\phi}=0.0871$ GeV. The $\phi BD$ coupling is fixed to the SU(3)-average value among the different decuplet-to-octet pionic decay channels, $\mathcal{C}=0.85$~\cite{Alarcon:2012nr}. The $\phi DD$ coupling $\mathcal{H}$ is barely known, and we fix it using the large $N_c$ relation $H_A=(9/5)g_A$, where
$g_A$ and $H_A$ are the nucleon and $\Delta$ axial charges.  With $g_A=1.26$, this yields the $\phi DD$ coupling $\mathcal{H}=H_A/2=1.13$. In the loop function~Eq.~\eqref{Eq:decmassG},  the LO corrections to the virtual octet masses are included; therefore, there are four more LECs $m_0$, $b_0$, $b_D$, and $b_F$ related to the octet baryon masses up to $\mathcal{O}(p^2)$. Similar to the determination of the decuplet baryon masses at $\mathcal{O}(p^2)$~\cite{Ren:2013dzt}, their values can be obtained by fitting the physical octet baryon masses with the NLO octet mass formula $M_B=m_0-m_B^{(2)}$. Because at the same pion masses, the $m_0$ and $b_0$ cannot be disentangled, we only obtain  $m_0^{\rm eff}=m_0-b_0(4M_K^2+2M_{\pi}^2)$, $b_D=0.06$ GeV$^{-1}$, and $b_F=-0.231$ GeV$^{-1}$. The octet-decuplet mass splitting $\delta=0.231$ GeV is taken as the average gap of the physical octet and decuplet masses. As a result,  $m_0$ and $b_0$ can be expressed as $m_0=m_D-0.231$ GeV and $b_0=(m_D-1.423)/1.014$ GeV$^{-1}$.

In the fitting process, we incorporate the inverse of the correlation matrix $C_{ij}=\sigma_i\sigma_j\delta_{ij} + \Delta a_i \Delta a_j$ for each lattice ensemble to calculate the $\chi^2$, where $\sigma_i$ are the lattice statistical errors and the $\Delta a_i$ are the fully correlated errors propagated from the determination of $a_i$. This is because the data from different collaborations are not correlated with each other, but the data from the same collaboration are partially correlated by the uncertainties propagated from the determination of the lattice spacing $a$.

\subsection{Light-quark mass dependence of the decuplet baryon masses}\label{subsec3b}

In this subsection, we proceed to study the eight sets of LQCD data for the decuplet baryon masses by using the N$^3$LO BChPT mass formulas [Eq.~\eqref{Eq:decmass2}]. In order to constrain better the values of the LECs, we include the precise experimental data in the fitting. The obtained $14$ LECs from the best fits are tabulated in Table~\ref{Tab:fitcoef}. For the sake of comparison, we also perform fits at  NLO~\footnote{Because at $\mathcal{O}(p^2)$ BChPT does not generate any FVCs, we have adjusted the lattice data by subtracting the FVCs calculated by the N$^3$LO EOMS BChPT.} and NNLO. Up to NNLO, there are only three  LECs, i.e., $m_D$, $t_0$, and $t_D$.

\begin{table}[t]
\centering
\caption{Values of the LECs from the best fits to the LQCD data and the experimental data with different fitting strategies at $\mathcal{O}(p^2)$, $\mathcal{O}(p^3)$, and $\mathcal{O}(p^4)$, respectively. The estimator for the fits with and without the experimental decuplet masses, $\chi^2/{\rm d.o.f.}$ and ${\chi^2/{\rm d.o.f.}}^*$, are given in the last two rows (see text for details).}
\label{Tab:fitcoef}
\begin{tabular}{lccc}
\hline\hline
                  &  NLO   &   NNLO & N$^3$LO      \\
\hline
  $m_D$~[MeV]         & $1135(14)$    & $870(12)$  &  $1152(25)$     \\
  $t_0$~[GeV$^{-1}$]  & $0.167(27)$   & $1.36(2)$  &  $0.0710(59)$      \\
  $t_D$~[GeV$^{-1}$]  & $0.322(2)$    & $0.785(3)$ &  $0.318(16)$       \\
  $\tilde{t}_1$~[GeV$^{-1}$]  & --    & -- &  $5.90(24)$          \\
  $\tilde{t}_2$~[GeV$^{-1}$]  & --    & -- &  $-2.26(29)$         \\
  $\tilde{t}_3$~[GeV$^{-1}$]  & --    & -- &  $-3.67(45)$         \\
  $t_7$~[GeV$^{-2}$]  & --            & -- &  $-2.37(8)$         \\
  $t_8$~[GeV$^{-2}$]  & --            & -- &  $0.298(156)$        \\
  $t_9$~[GeV$^{-2}$]  & --            & -- &  $1.21(13)$         \\
  $e_1$~[GeV$^{-3}$]  & --            & -- &  $-0.00386(11689)$     \\
  $e_2$~[GeV$^{-3}$]  & --            & -- &  $0.194(47)$       \\
  $e_3$~[GeV$^{-3}$]  & --            & -- &  $-0.167(117)$        \\
  $e_4$~[GeV$^{-3}$]  & --            & -- &  $0.0767(480)$        \\
  $e_5$~[GeV$^{-3}$]  & --            & -- &  $-0.0182(734)$           \\
  \hline
$\chi^2/{\rm d.o.f.}$ & $4.4$ &  $9.5$ & $0.20$         \\
  \hline
${\chi^2/{\rm d.o.f.}}^*$ & $0.44$ &  $1.7$ & $0.18$         \\
\hline\hline

\end{tabular}

\end{table}

It is clear that the NLO fit (without loop contributions) already describes the LQCD simulations very well. The description becomes
a bit worse at NNLO.\footnote{Without the contributions of the virtual octet baryons, the NNLO description would be much better, with
a $\chi^2/{\rm d.o.f.}\approx 2.9$.} While the description at N$^3$LO becomes much better, yielding a $\chi^2/{\rm d.o.f.}=0.20$. Therefore we confirm that the PACS-CS, QCDSF-UKQCD, and HSC data are consistent with each other, although their setups are different.  Furthermore, it seems that the LQCD decuplet baryon masses are almost linear in $M_{\pi}^2$, as demonstrated by the  good fit obtained at NLO, ${\chi^2/{\rm d.o.f.}}^*=0.44$.

 The values of the $14$ LECs seem very natural, except that the LECs $\tilde{t}_1$, $\tilde{t}_2$, $\tilde{t}_3$, and $t_7$ might be
 slightly large. If we had constrained their values  to lie between $-1$ to $1$ in the fitting process, we would have obtained a $\chi^2/\mathrm{d.o.f.}= 0.23$, instead of $0.20$. It is evident that the present LQCD simulations are not precise enough or are too limited to put a stringent constraint on the values of all the LECs appearing up to N$^3$LO, because the NLO fit already yields a $\chi^2/\mathrm{d.o.f.}^*$ smaller than 1. This is further
 confirmed by the relatively large correlation observed between some of the LECs, e.g.,  between $\tilde{t}_1$ and $\tilde{t}_2$, among $t_7$, $t_8$, and $t_9$, and among $e_1$, $e_3$, and $e_5$. We found that putting some of these LECs to zero only slightly increases the $\chi^2/\mathrm{d.o.f.}$. In short, the values of the N$^3$LO LECs and the corresponding uncertainties should be viewed in the present context and used with care.

As mentioned earlier, the lightest LQCD point with $M_\pi=296$ MeV of the PACS-CS Collaboration suffers from potentially large systematic errors. If we had performed the fit without this point, we would have obtained  a $\chi^2/\mathrm{d.o.f.}=0.24$, slightly larger than the $\chi^2/\mathrm{d.o.f.}=0.20$ of Table \ref{Tab:fitcoef}. In addition,  the values of the corresponding LECs would change moderately. On the other hand, the extrapolations with the LECs determined from the fit excluding the physical masses became much worse. This seems to suggest that the inclusion of the lightest PACS-CS point  is reasonable, keeping in mind the caveat that they may suffer from potentially large systematic errors. This is also the strategy adopted by the PACS-CS Collaboration~\cite{Ishikawa:2009vc} and other similar studies~\cite{Semke:2012gs}.

\begin{figure}[t]
  \centering
  \includegraphics[width=12cm]{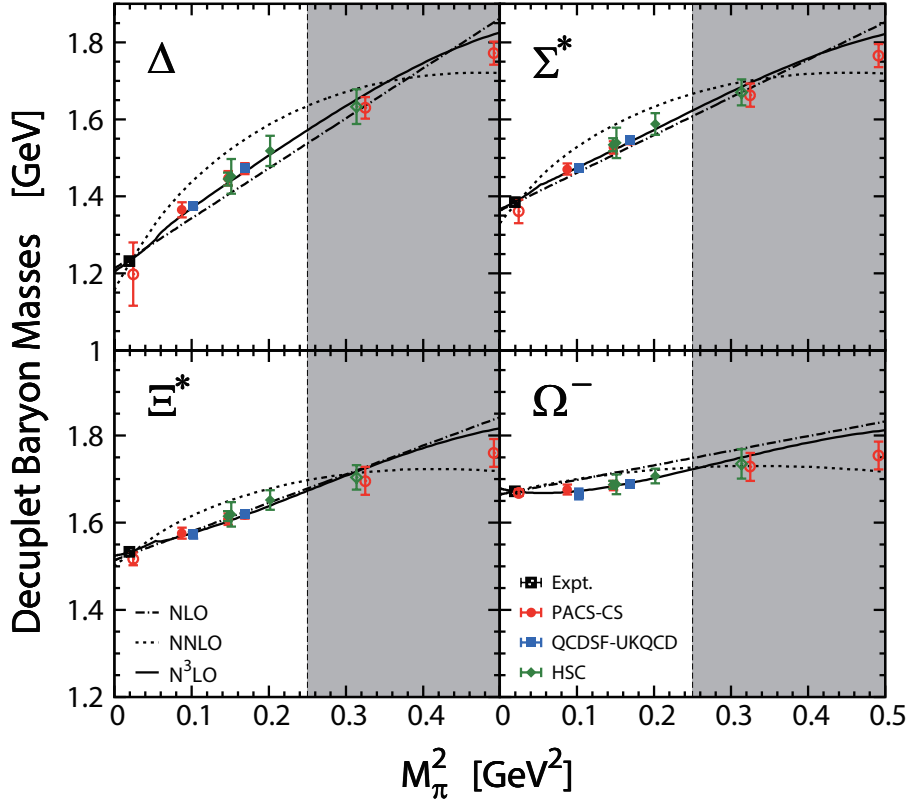}\\
  \caption{(Color online). Pion mass dependence of the lowest-lying decuplet baryon masses. Filled (open) symbols denote the lattice data points included in (excluded from) the fits, which are projected to have the physical strange-quark mass. The dot-dashed, the dashed, and the solid lines are  the best NLO, NNLO and N$^3$LO fits to the lattice data, respectively. In obtaining the BChPT results, the strange-quark mass has been set to its physical value. The lattice points in the shaded region are
  not included in the fits.}\label{Fig:chiral-1}
\end{figure}

In Fig.~\ref{Fig:chiral-1}, we show the $\Delta$, $\Sigma^*$, $\Xi^*$, and $\Omega^-$ masses as functions of $M_{\pi}^2$, where the strange-quark mass is set at its physical value. It is clear that the LQCD data are rather linear in $M_{\pi}^2$ . The $\mathcal{O}(p^3)$ BChPT results show strong curvature and cannot describe the LQCD data. A good description can only be achieved up to N$^3$LO.\footnote{In principle, at NNLO,  we can use for
the meson-decay constant its SU(3) average, $F_\phi=1.17f_\pi$ with $f_\pi=92.4$ MeV. This improves a lot the NNLO  fit.} In Fig.~2, we also show those data of the PACS-CS and HSC Collaborations that are excluded from the fit. The $\mathcal{O}(p^4)$ BChPT can describe reasonably well those data as well.

It should be emphasized that the setups of the QCDSF-UKQCD simulations are rather different from those of  the PACS-CS and HSC Collaborations. Most LQCD simulations fix the strange-quark mass at (or close to ) its physical value and gradually moving the $u/d$ quark masses to their physical values. The QCDSF-UKQCD Collaboration adopted an alternative method by starting at a point on the SU(3) flavor symmetric line ($m_{u/d}=m_s$) and holding the sum of the quark masses $\bar{m}=(2m_{u/d}+m_s)/3$ constant~\cite{Bietenholz:2010jr}. In this way, the corresponding kaon and eta masses  can be smaller than the pion mass. On the other hand, the FVCs from the kaon and eta loops can become comparable or even larger than that induced by the pion loop, because the $M_{\phi}L$ can simultaneously  become smaller than $4$. Therefore, the QCDSF-UKQCD data provide us an opportunity to test the BChPT in the world of small strange-quark masses and small lattice volumes.

\begin{figure}[t]
  \centering
  \includegraphics[width=10cm]{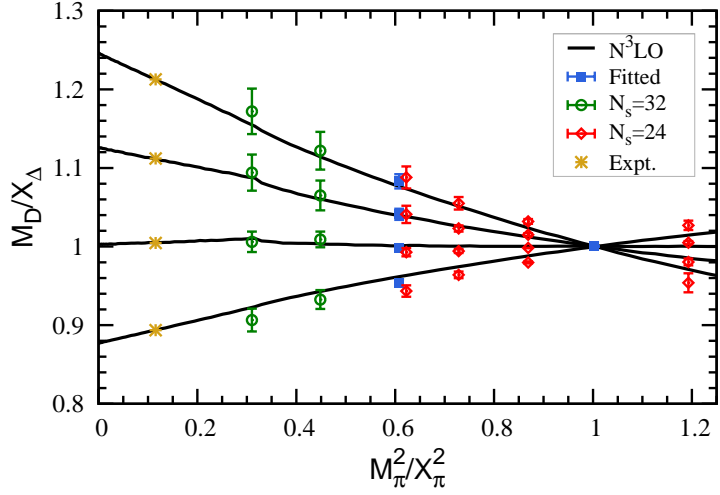}\\
  \caption{(Color online). The QCDSF-UKQCD lattice data~\cite{Bietenholz:2010jr} in comparison with the N$^3$LO BChPT. The lattice data denoted by the blue filled squares  are included in the fit; those by the green opened circles (with $N_s=32$) and the red diamonds (with $N_s=24$) are not. FVCs of the lattice results have been subtracted.  The two-flavour singlet quantities, $X_\pi$ and $X_\Delta$, are
  defined as, $X_{\pi}=\sqrt{(M_{\pi}^2+2M_K^2)/3}$, $X_{\Delta}=(2m_{\Delta}+m_{\Omega^-})/3$, respectively~\cite{Bietenholz:2010jr}.}
  \label{Fig:desUKQCD}
\end{figure}

In Fig.~\ref{Fig:desUKQCD}, the QCDSF-UKQCD  data are compared with the N$^3$LO BChPT. The LQCD points included in the fit are denoted by solid points and those excluded from the fit by hollow points. All lattice points are  shifted by FVCs and the kaon mass is fixed using the function $M_{K}^2=a+bM_{\pi}^2$ for the lattice ensemble with $a$ and $b$ determined in Appendix II of Ref.~\cite{Ren:2012aj}. It is clear that the N$^3$LO BChPT can describe reasonably well the QCDSF-UKQCD data obtained in both large ($N_s=32$) and small ($N_s=24$) volumes with both heavy and light pion masses. However, it should be pointed out that the ratio method eliminates to a large extent the FVCs. In other words, to plot/study the data this way one can neglect FVCs, as noticed in Ref.~\cite{Bietenholz:2011qq}.

In Table \ref{Tab:FVCs}, we show the FVCs to the LQCD data calculated in the N$^3$LO BChPT. Most of them are at the order of a few of tens of MeV. Among them, the FVCs to the QCDSF-UKQCD data are the largest, which can be easily understood from the arguments given above.
\begin{table}[h!]
  \centering
  \caption{Finite-volume corrections (in units of MeV) to LQCD decuplet baryon masses in covariant BChPT up to N$^3$LO.}
  \label{Tab:FVCs}
  \begin{tabular}{c|cc|cccc|cccc}
    \hline\hline
      & $M_\pi$ & $M_K$ & $\delta m_\Delta$ & $\delta m_{\Sigma^*}$ & $\delta m_{\Xi^*}$ & $\delta m_{\Omega^-}$ & $M_\pi L$ & $M_K L$ & $M_\eta L$ \\
    \hline
PACS-CS & 296  & 594 & 14 & 5 & 0 & $-3$ & 4.3 & 8.7 & 9.8 \\
        & 384  & 581 & 5 & 2  & 1  & 1 & 5.7 & 8.6 & 9.3  \\
        & 411  & 635 & 4 & 2  & 0  & 1 & 6.0 & 9.3 & 10.2 \\
    \hline
QCDSF-UKQCD & 320 & 451 & 20 & 13 & 8 & 4 & 4.1 & 5.8 & 6.2 \\
            & 411 & 411 & 50 & 50 & 50 & 50 & 3.95 & 3.95 & 3.95 \\
    \hline
    HSC & 383 & 544 & 4  & 2  & 1  & 0 & 5.7 & 8.1 & 8.8 \\
        & 389 & 546 & 42 & 27 & 14 & 3 & 3.9 & 5.4 & 5.9 \\
        & 449 & 581 & 28 & 19 & 11 & 4 & 4.5 & 5.8 & 6.2 \\
    \hline\hline
  \end{tabular}
\end{table}

\begin{figure}[t]
  \centering
  \includegraphics[width=10cm]{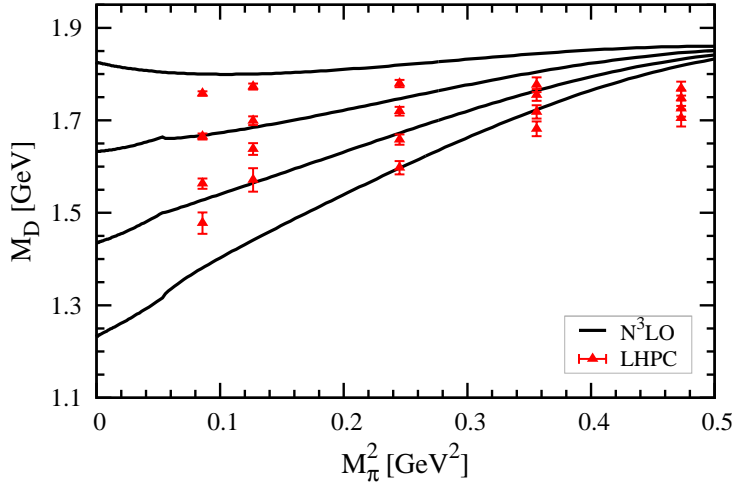}\\
  \caption{(Color online). Comparison between the N$^3$LO BChPT  and the LHPC data~\cite{WalkerLoud:2008bp}.}
  \label{Fig:desLHPC}
\end{figure}

We would like to point out that in the above fits we have not included the LHPC data, while in Refs.~\cite{Ren:2012aj,Ren:2013dzt} we have studied
their data for the octet baryon masses. The reason is  that the LHPC decuplet baryon data
do not seem to be consistent with those of the PACS-CS, QCDSF-UKQCD, and HSC  Collaborations.
This is clearly demonstrated in Fig.~\ref{Fig:desLHPC}, where the LHPC data are contrasted with the N$^3$LO BChPT with the N$^3$LO LECs
tabulated in Table \ref{Tab:fitcoef}, and the corresponding kaon mass is fixed using $M_{K}^2=a+bM_{\pi}^2$ with $a$ and $b$ determined in Ref.~\cite{Ren:2012aj}.
It is clear that the dependencies of the lattice data on $M_\pi^2$ seem to be flatter than suggested by the N$^3$LO BChPT.
In Ref.~\cite{WalkerLoud:2008bp}, it was noticed that it is difficult to extrapolate the LQCD data to the physical $\Delta(1232)$  mass. Our study seems to confirm their finding. If we had included the LHPC data\footnote{It needs to be mentioned that
in Ref.~\cite{WalkerLoud:2011ab}, a different way of setting the lattice scale has been used to obtain the decuplet baryon masses of the LHPC Collaboration~\cite{WalkerLoud:2008bp}
in physical units.}(three sets of them satisfying our selection criteria) in our fitting, we would have obtained
a $\chi^2/{\rm d.o.f.}=2.2$.

Furthermore, in order to quantify the effects of loop contributions involving virtual octet and decuplet baryons,
one can allow $\mathcal{C}$ and $\mathcal{H}$ to vary in the fitting. The corresponding $\chi^2/{\rm d.o.f.}$ from the best fit is 0.23 with $\mathcal{C}=0.75$ and $\mathcal{H}=1.0$. It is clear that the values are consistent with the phenomenological values we used above, which can be seen as
evidence for the existence of non-analytical chiral contributions following the argument given in Ref.~\cite{WalkerLoud:2011ab}. One should note, however, that because of the small difference between the $\chi^2/\mathrm{d.o.f.}$ obtained here and the $\chi^2/\mathrm{d.o.f.}$ obtained by putting $\mathcal{C}$ and $\mathcal{H}$ to zero, this evidence is rather weak in the present case.

\subsection{Convergence of SU(3) EOMS BChPT}

Convergence of BChPT in the $u$, $d$, and $s$ three-flavor sector has been under debate for many years. See, e.g., Refs.\cite{Donoghue:1998bs,Bernard:2003rp,Young:2002ib,Geng:2013xn} and references cited therein.\footnote{For related discussions in
the mesonic sector, see, e.g., Refs.~\cite{DescotesGenon:2003cg,Bernard:2010ex}, where the so-called resummed chiral perturbation theory has been shown to exhibit better convergence than conventional chiral perturbation theory. To our knowledge, no similar studies exist in the one-baryon sector.} One prominent example is the magnetic moments of octet baryons. In Ref.~\cite{Geng:2008mf}, it has been shown that compared to the HB ChPT and the IR BChPT,
the EOMS BChPT converges relatively faster. The same has been found for the octet baryon masses~\cite{MartinCamalich:2010fp}. Nevertheless, even in the EOMS BChPT,
convergence is relatively slow because of the large expansion parameter, $M_K/\Lambda_\mathrm{ChPT}$. Naively, each higher-order contribution is only suppressed by about one-half at the physical point, which can even be further reduced for LQCD simulations with larger light-quark masses.  To speed up convergence, several alternative formulations of BChPT have been proposed, such as the long distance regularization method~\cite{Donoghue:1998bs}, the cutoff scheme~\cite{Bernard:2003rp}, and finite-range regulator method~\cite{Young:2002ib,Leinweber:2003dg} BChPT, which exhibit better convergence by suppressing loop contributions with either a cutoff or a form factor.

\begin{table}[b]
  \centering
  \caption{Contributions of different chiral orders  to the decuplet baryon masses at the physical point (in units of GeV).}
  \label{Tab:order}
\begin{tabular}{l|c|ccc|ccc|ccc|ccc}
  \hline\hline
      &  & \multicolumn{3}{c|}{$\Delta$} & \multicolumn{3}{c|}{$\Sigma^*$}  & \multicolumn{3}{c|}{$\Xi^*$} & \multicolumn{3}{c}{$\Omega^-$}\\
     \cline{3-5} \cline{6-8} \cline{9-11}  \cline{12-14}
     & $m_D$ & $p^2$ & $p^3$ & $p^4$  & $p^2$ & $p^3$ & $p^4$ &  $p^2$ & $p^3$ & $p^4$  &  $p^2$ & $p^3$ & $p^4$\\
  \hline
  NLO & $1.135$  & $0.104$ & -- & -- & $0.248$ & -- & --  & $0.392$ & -- & -- & $0.537$ & -- & -- \\
  NNLO & $0.870$  & $0.737$ & $-0.383$ & -- & $1.089$ & $-0.582$ & -- & $1.441$ & $-0.785$ & -- & $1.793$ & $-0.991$ & -- \\
  N$^3$LO & $1.152$ & $0.046$ & $-0.429$ & $0.463$  & $0.158$ & $-0.652$ & $0.728$  & $0.270$ & $-0.878$ & $0.988$  & $0.382$ & $-1.106$ & $1.244$ \\
  \hline\hline
\end{tabular}
\end{table}

\begin{figure}[h]
  \centering
  \includegraphics[width=12cm]{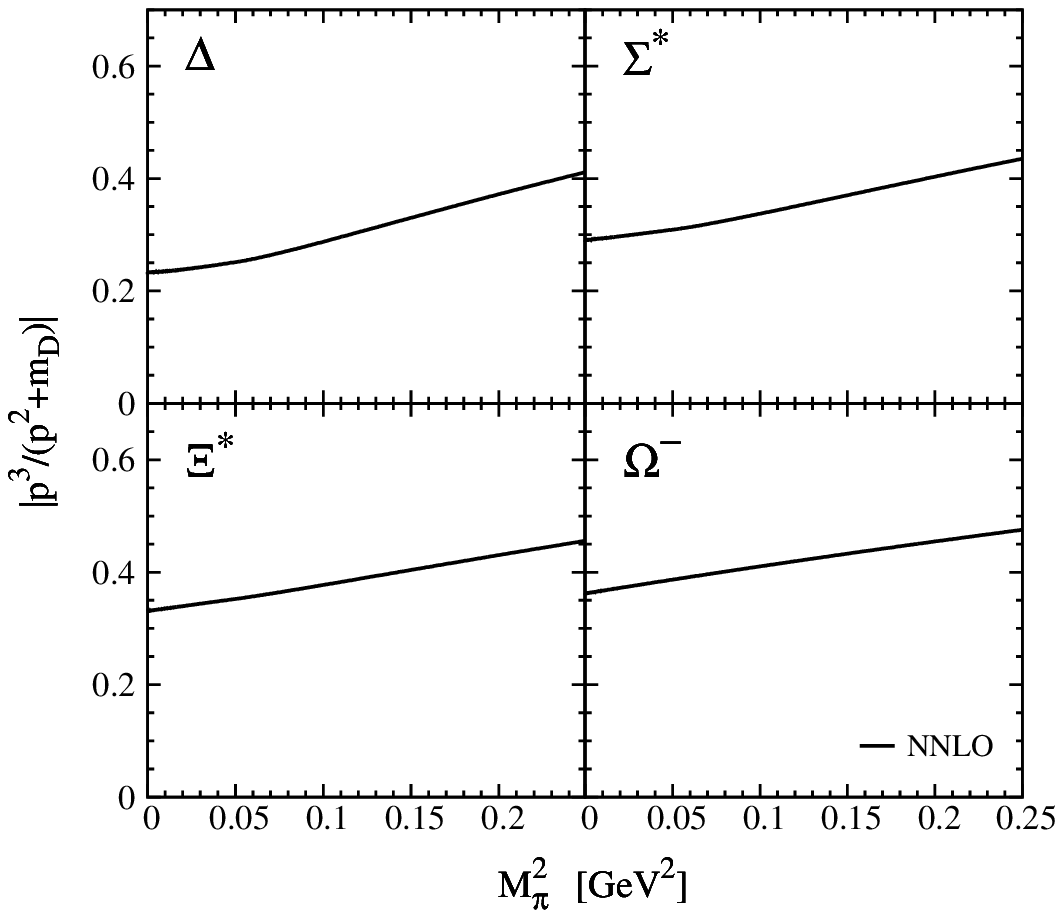}\\
  \caption{Ratio of one-loop and tree contributions to the decuplet baryon masses, $\left|p^3/(p^2+m_D)\right|$, as a function of pion mass. The strange-quark mass is set at its physical value.}
  \label{Fig:p34p2}
\end{figure}
\begin{figure}[h]
  \centering
  \includegraphics[width=12cm]{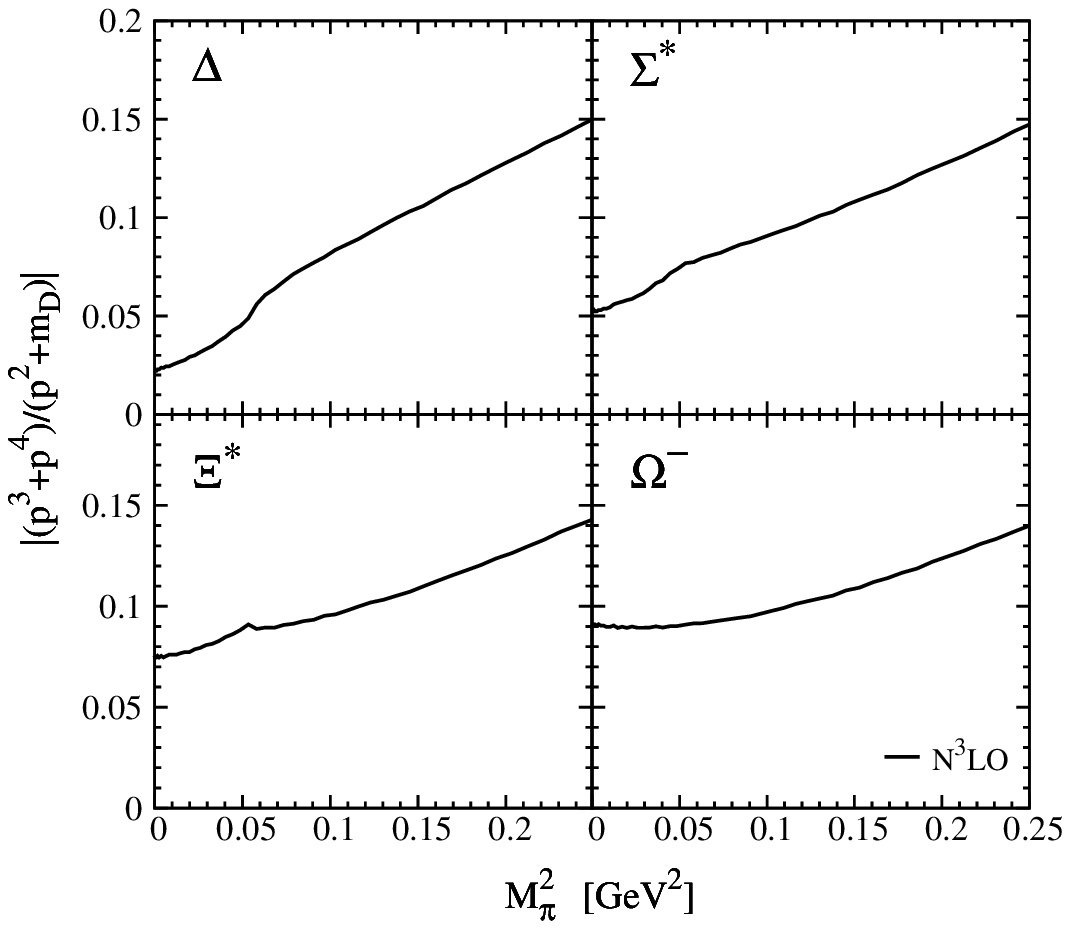}\\
  \caption{Ratio of one-loop and tree contributions to the decuplet baryon masses, $\left|(p^3+p^4)/(p^2+m_D)\right|$, as a function of pion mass. The strange-quark mass is set at its physical value.}
  \label{Fig:p34p2md}
\end{figure}
In the following, we would like to examine the contributions of different chiral orders. In Table~\ref{Tab:fitcoef} and Fig.~\ref{Fig:chiral-1}, one notices that the NLO BChPT can already describe the LQCD data very well, but the experimental data are missed a little bit. Naturally one would expect that up to NNLO and  N$^3$LO, there should be some reshuffling of contributions of different orders.  This can be clearly seen from Table \ref{Tab:order}, where contributions of different chiral orders to the decuplet baryon masses at the physical point are tabulated. On the other hand, once loop diagrams are included, a naive comparison of $p^0$ ($m_D$), $p^2$, $p^3$, and $p^4$ contributions turns out to be troubling. At NNLO, the $p^2$ contributions can be a factor of 2 larger than $m_D$, while at N$^3$LO, the $p^3$ and $p^4$ contributions are opposite and become comparable to or even larger than the $p^2$ contributions, particularly for the decuplet baryons containing strangeness.

On the other hand, up to one-loop level, it might be more proper to judge convergence by comparing
tree-level and loop contributions. In Figs. \ref{Fig:p34p2} and \ref{Fig:p34p2md}, $\left|p^3/(p^2+m_D)\right|$ and $\left|(p^3+p^4)/(p^2+m_D)\right|$ are shown as a function of $M_\pi^2$. At NNLO, the $p^3$ contributions can reach about  50\% of  the tree-level  contributions, while at N$^3$LO the loop  contributions become about $10\%\sim20\%$ of the tree-level contributions. These results suggest that the chiral expansions are convergent as expected.

\subsection{Pion- and strangeness-baryon sigma terms}
The baryon sigma terms are important quantities in understanding the chiral condensate and the composition of the baryons. At present, there is no direct  LQCD simulation of these quantities for the decuplet baryons.  On the other hand,
 one can calculate the decuplet baryon sigma terms $\sigma_{\pi D}$ and $\sigma_{s D}$ using
 BChPT, once the relevant LECs are fixed, via the Feynman-Hellmann theorem  by treating the decuplet baryons as stable particles as in standard BChPT.  See, e.g., Ref.~\cite{Ren:2012aj} for
relevant formulas.

\begin{table}[t]
  \centering
  \caption{Pion- and strangeness-sigma terms of the decuplet baryons at the physical point. The first error is statistical and the second is systematic, estimated by taking half the difference between the N$^3$LO result and the NNLO result.}
  \label{Tab:sigma}
  \begin{tabular}{c|cc|cc}
    \hline\hline
     & \multicolumn{2}{c|}{NNLO}  & \multicolumn{2}{c}{N$^3$LO}\\
     \cline{2-3} \cline{4-5}
     & This work & Ref.~\cite{MartinCamalich:2010fp}  & This work & Ref.~\cite{Semke:2012gs} \\
    \hline
    $\sigma_{\pi \Delta}$ [MeV]    & $64(1)$ & $55(4)(18)$ & $28(1)(18)$ & $34(3)$ \\
    $\sigma_{\pi \Sigma^*}$ [MeV]  & $44(1)$ & $39(3)(13)$ & $22(2)(11)$ & $28(2)$\\
    $\sigma_{\pi \Xi^*}$ [MeV]     & $26(1)$ & $22(3)(7)$  & $11(2)(8)$ & $18(4)$\\
    $\sigma_{\pi \Omega^-}$ [MeV]  & $8(1)$  & $5(2)(1)$   & $-5(2)(6)$  & $10(4)$ \\
    \hline
    $\sigma_{s \Delta}$ [MeV]      & $93(12)$  & $56(24)(1)$    & $88(22)(3)$   &  $41(41)$ \\
    $\sigma_{s \Sigma^*}$ [MeV]    & $181(13)$ & $160(28)(7)$   & $243(24)(31)$ & $211(44)$\\
    $\sigma_{s \Xi^*}$ [MeV]       & $258(14)$ & $274(32)(9)$   & $391(24)(67)$ & $373(53)$\\
    $\sigma_{s \Omega^-}$ [MeV]    & $326(15)$ & $360(34)(26)$  & $528(26)(101)$& $510(50)$ \\
    \hline\hline
  \end{tabular}
\end{table}

Using the LECs given in Table~\ref{Tab:fitcoef}, we calculate the sigma terms of the baryon decuplet at the physical point,
and the  results are listed in Table~\ref{Tab:sigma}. For comparison, we also tabulate the results of Refs.~\cite{MartinCamalich:2010fp,Semke:2012gs}.  The difference between our $\mathcal{O}(p^3)$ predictions with those
of Ref.~\cite{MartinCamalich:2010fp} reflects the influence of the LQCD data and the fitting strategy.  While our N$^3$LO results are
consistent with those of Ref.~\cite{Semke:2012gs} within uncertainties.

\section{Summary}\label{SecIV}

We have studied the ground-state decuplet baryon masses in  baryon chiral perturbation theory with the extended-on-mass-shell scheme up to
next-to-next-to-next-to-leading order.
Through a simultaneous fit of the $n_f=2+1$ LQCD data from the PACS-CS, QCDSF-UKQCD, and HSC Collaborations, the $14$ unknown low-energy constants are determined. In fitting the LQCD data, finite-volume corrections are taken into account self-consistently.  A $\chi^2/{\rm d.o.f.}=0.20$
is achieved for the eight sets of  LQCD data satisfying $M_{\pi}^2<0.25$ GeV$^2$ and $M_{\phi}L>3.8$.

 Our studies show that the chiral expansions are convergent as expected and the results of the PACS-CS, QCDSF-UKQCD, and HSC Collaborations seem to be
 consistent with each other, but not those of the LHPC Collaboration. We have calculated the sigma terms of the decuplet baryons by use of the Feynman-Hellmann theorem, which should be compared to future LQCD data.

It should be noted that our present study suffers from the limited range of the LQCD data (in terms of the input parameters) and the rather large number of unknown low-energy constants.
Future refined LQCD simulations with various light-quark and strange-quark masses, lattice volume and lattice spacing will be extremely welcome to put covariant baryon chiral perturbation theory to a more stringent test than was done in the present work. In the context of effective field theories, one would like to apply the same formalism and utilize the same low-energy constants to study other related physical observables,
which can also serve as an additional test. Such works are in progress and will be reported elsewhere.

\begin{acknowledgments}
X.-L.R  thanks Dr. Hua-Xing Chen for useful discussions and acknowledges support from the Innovation
Foundation of Beihang University for Ph.D. Graduates.
L.-S.G acknowledges support from the Alexander von Humboldt foundation.
This work was partly supported by the National
Natural Science Foundation of China under Grants No. 11005007, No. 11035007, and No. 11175002,  the New Century Excellent Talents in the University  Program of Ministry of Education of China under Grant No. NCET-10-0029,  the Fundamental Research Funds for the Central Universities,
and the Research Fund  for the Doctoral Program of Higher Education under Grant No. 20110001110087.

\end{acknowledgments}

\section{Appendix}
Here we show explicitly the N$^3$LO loop functions appearing in Eq.~(\ref{Eq:decmass2}), which are calculated in the EOMS scheme:
\begin{eqnarray}
  H_{D}^{(e,1)}(M_{\phi}) &=& M_{\phi}^2\left[1+{\rm ln}\left(\frac{\mu^2}{M_{\phi}^2}\right)\right],\\
  H_{D}^{(e,2)}(M_{\phi}) &=& M_{\phi}^4\left[1+{\rm ln}\left(\frac{\mu^2}{M_{\phi}^2}\right)\right],\\
  H_{D}^{(e,3)}(M_{\phi}) &=& m_D\left\{\frac{M_{\phi}^4}{4}\left[1+{\rm ln}\left(\frac{\mu^2}{M_{\phi}^2}\right)\right] +\frac{1}{8}M_{\phi}^4\right\}.
\end{eqnarray}
\begin{eqnarray}\label{eq:w}
   H_{D,B}^{(f)}(M_{\phi}) &=& \frac{1}{144m_D^4}m_D^{(2)}M_{\phi}^2 \left[90m_0^4+96m_0^3m_D+36m_0^2m_D^2+48m_0m_D^3-22m_D^4\right.\nonumber\\
   &&\quad \left. -3\left(30m_0^2+16m_0m_D+19m_D^2\right)M_{\phi}^2+30M_{\phi}^4\right]\nonumber\\
   && - \frac{1}{12m_D^3}m_B^{(2)}M_{\phi}^2 \left[8m_0^3+8m_0^2m_D+4m_0m_D^2+7m_D^3-(4m_0+m_D)M_{\phi}^2\right]\nonumber\\
   && + \frac{1}{24m_D^6}m_D^{(2)}M_{\phi}^2\ln\left(\frac{M_{\phi}}{m_0}\right) \left[4m_0^5(5m_0+6m_D)\right.\nonumber\\
   && \left.\quad- 6(5m_0^4+4m_0^3m_D+3m_0^2m_D^2+2m_0m_D^3+m_D^4)M_{\phi}^2 + 4(5m_0^2+2m_0m_D+3m_D^2)M_{\phi}^4-5M_{\phi}^6\right]\nonumber\\
   && -\frac{1}{12m_D^5}m_B^{(2)}M_{\phi}^2\ln\left(\frac{M_{\phi}}{m_0}\right) \left[3m_0^4(4m_0+5m_D)\right.\nonumber\\
   && \left.\quad -3(4m_0^3+3m_0^2m_D+2m_0m_D^2+m_D^3)M_{\phi}^2+(4m_0+m_D)M_{\phi}^4\right]\nonumber\\
   && +\frac{1}{24m_D^6}(m_0-m_D)^2(m_0+m_D)^4\left[\ln(m_0M_{\phi})-\ln(m_0^2-m_D^2)-i\pi\right]\nonumber\\
   && \quad \times\left[2m_Dm_B^{(2)}(4m_0-m_D)-m_D^{(2)}(5m_0^2-2m_0m_D+3m_D^2)\right]\nonumber\\
   && + \frac{1}{12}M_{\phi}^2\left(3m_B^{(2)}+2m_D^{(2)}\right) \ln\left(\frac{m_0M_{\phi}}{\mu^2}\right)\nonumber\\
   && +\frac{1}{24m_D^6\sqrt{\mathcal{W}}} (m_0^2-2m_0m_D+m_D^2-M_{\phi}^2)(m_0^2+2m_0m_D+m_D^2-M_{\phi}^2)^2\nonumber\\
   &&\quad\times \left[5m_0^4m_D^{(2)}-2m_0^2 \left(5M_{\phi}^2m_D^{(2)}+m_D^2\left(m_D^{(2)}-m_B^{(2)}\right)\right) -2m_0^3m_D\left(m_D^{(2)}+4m_B^{(2)}\right)\right.\nonumber\\
   && \quad\quad \left. +2m_0m_D(m_D^2+M_{\phi}^2)\left(m_D^{(2)}+4m_B^{(2)}\right) - (m_D^2-M_{\phi}^2)\left(5M_{\phi}^2m_D^{(2)}+m_D^2\left(3m_D^{(2)}+2m_B^{(2)}\right)\right)\right]\nonumber\\
   &&\quad\times \left[\arctan\left(\frac{m_0^2-m_D^2-M_{\phi}^2}{\sqrt{\mathcal{W}}}\right) -\arctan\left(\frac{m_0^2+m_D^2-M_{\phi}^2}{\sqrt{\mathcal{W}}}\right)\right],
\end{eqnarray}
\begin{eqnarray}\label{eq:g}
  H_{D,D^{\prime}}^{(g)}(M_{\phi}) &=& \frac{M_{\phi}^2}{432m_D^4}\left[4m_D^4\left(132m_D^{(2)}- 97m_{D^{\prime}}^{(2)}\right)\right.\nonumber\\
  && \quad \left.+30M_{\phi}^4\left(3m_D^{(2)}+2m_{D^{\prime}}^{(2)}\right)
  -15m_D^2M_{\phi}^2\left(31m_D^{(2)}+ 14m_{D^{\prime}}^{(2)}\right)\right]\nonumber\\
  && +\frac{5M_{\phi}^2}{72m_D^6}\ln\left(\frac{M_{\phi}}{m_D}\right)\left[-30m_D^4M_{\phi}^2m_D^{(2)}+
  48m_D^6\left(m_D^{(2)}-m_{D^{\prime}}^{(2)}\right)\right.\nonumber\\
  && \quad \left. + 10m_D^2M_{\phi}^4\left(2m_D^{(2)}+m_{D^{\prime}}^{(2)}\right)
  -M_{\phi}^6\left(3m_D^{(2)}+2m_{D^{\prime}}^{(2)}\right)\right]\nonumber\\
  && + \frac{5M_{\phi}^2}{36}\left(12m_D^{(2)}-7m_{D^{\prime}}^{(2)}\right)
  \ln\left(\frac{m_D^2}{\mu^2}\right)\nonumber\\
  && -\frac{5}{72m_D^6}M_{\phi}^3(4m_D^2-M_{\phi}^2)^{3/2}
  \left[2m_D^2\left(m_D^{(2)}-m_{D^{\prime}}^{(2)}\right) -M_{\phi}^2\left(3m_D^{(2)}+2m_{D^{\prime}}^{(2)}\right)\right]\nonumber\\
  &&\quad \times \left[\arctan\left(\frac{M_{\phi}}{\sqrt{4m_D^2-M_{\phi}^2}}\right) +\arctan\left(\frac{2m_D^2-M_{\phi}^2}{M_{\phi}\sqrt{4m_D^2-M_{\phi}^2}}\right)\right].
\end{eqnarray}
In Eqs.~(\ref{eq:w}) and (\ref{eq:g}), $\mathcal{W}=-m_0^4-(m_D^2-M_{\phi}^2)^2+2m_0^2(m_D^2+M_{\phi}^2)$,
 $m_D^{(2)}$ and $m_B^{(2)}$ are the NLO decuplet and octet baryon masses, where
 $m_D^{(2)}$ is given in Eq.~\eqref{Eq:decmass2}, and  $m_B^{(2)}$ has the following form:
\begin{equation}\label{Eq:BLO}
  m_B^{(2)}=\sum\limits_{\phi=\pi,~K}\xi_{B,\phi}^{(2)} M_{\phi}^2
\end{equation}
with the corresponding coefficients $\xi_{B,\phi}^{(2)}$ listed in Table~\ref{Tab:NLO-octet}.
\begin{table}[t]
\centering
\caption{Coefficients of the NLO contributions to the octet baryon masses [Eq.~\eqref{Eq:BLO}].}
\label{Tab:NLO-octet}
\begin{tabular}{ccccc}
\hline\hline
    & $N$ & $\Lambda$ & $\Sigma$ & $\Xi$ \\
  \hline
  $\xi^{(a)}_{B,\pi}$ & $-(2b_0+4b_F)$ & $\frac{-2}{3}(3b_0-2b_D)$ & $-(2b_0+4b_D)$ & $-(2b_0-4b_F)$ \\
  $\xi^{(a)}_{B, K}$ & $-(4b_0+4b_D-4b_F)$ & $\frac{-2}{3}(6b_0+8b_D)$ & $-4b_0$ & $-(4b_0+4b_D+4b_F)$ \\
\hline\hline
\end{tabular}
\end{table}

\bibliographystyle{apsrev4-1}
\bibliography{decuplet-mass}

\end{document}